# How Complex Is a Fractal? Head/tail Breaks and Fractional Hierarchy

Bin Jiang and Ding Ma

Faculty of Engineering and Sustainable Development, Division of GIScience
University of Gävle, SE-801 76 Gävle, Sweden
Email: bin.jiang|ding.ma@hig.se

*(Draft: December 2016, Revision: January, April, June, August, and October 2017)*

**Abstract**
A fractal bears a complex structure that is reflected in a scaling hierarchy, indicating that there are far more small things than large ones. This scaling hierarchy can be effectively derived using head/tail breaks – a clustering and visualization tool for data with a heavy-tailed distribution – and quantified by an ht-index, indicating the number of clusters or hierarchical levels, a head/tail breaks-induced integer. However, this integral ht-index has been found to be less precise for many fractals at their different phrases of development. This paper refines the ht-index as a fraction to measure the scaling hierarchy of a fractal more precisely within a coherent whole, and further assigns a fractional ht-index – the fht-index – to an individual data value of a data series that represents the fractal. We developed two case studies to demonstrate the advantages of the fht-index, in comparison with the ht-index. We found that the fractional ht-index or fractional hierarchy in general can help characterize a fractal set or pattern in a much more precise manner. The index may help create intermediate map scales between two consecutive map scales.

## I. Introduction

All fractals bear a complex structure with far more small things than large ones. This notion of far more small things than large ones, being recursive in nature, can be expressed as a scaling hierarchy of numerous smallest things, a very few largest, and some in between the smallest and the largest. The scaling hierarchy can be revealed by head/tail breaks, which is a clustering and visualization tool for data with a heavy-tailed distribution (Jiang 2013, Jiang 2015a). More specifically, a data series is ranked from the largest to smallest, and then its average is to partition the data series into two parts: those greater than the average as the head, accounting for the minority of the data series, and those less than the average as the tail, accounting for the majority. This partition process continues for the head iteratively until the head is no longer a minority (for example, 40 percent). This recursive partition or head/tail breaks process leads to a number of clusters or hierarchical levels that are measured by the ht-index (Jiang and Yin 2014). To illustrate, given ten numbers that follow Zipf's law (Zipf 1949) exactly: 1, 1/2, 1/3, …, 1/10, their average is 0.29, which partitions the ten numbers into the largest three in the head and the remaining seven in the tail. For the three in the head 1, 1/2, and 1/3, their average is 0.61, which partitions the largest three in the head into one (1) in the head again, and two (1, 1/2) in the tail. Thus, the scaling pattern of far more small numbers than large ones recurs twice, so the ht-index of the ten numbers is three.

The head/tail breaks or ht-index has been used to re-define fractal, leading to the so-called third definition of fractal: *a set or pattern is fractal if the scaling pattern of far more small things than large ones recurs multiple times or with the ht-index being at least three* (Jiang and Yin 2014, Jiang 2015a). Under the new definition, a fractal is simply characterized by a data series of a heavy-tailed distribution, with the ht-index indicating its scaling hierarchy or complexity – the higher the ht-index of a fractal, the more complex the fractal. Currently, the ht-index is an integer, so complexity or scaling hierarchy is measured by whole numbers. In this paper, we will develop a real number to measure the scaling hierarchy in a more precise manner. There may be a variety of applications related to spatial data, including generalization.

**II. Motivation**
Scaling hierarchy cannot always be an integer. For example, a data series of 10 numbers {1, 1/2, 1/3, …, 1/10} has the ht-index of three, as seen above. If we append five small numbers such as 1/11, 1/12, …, and 1/15 into the data series to become {1, 1/2, 1/3, …, 1/10, 1/11, 1/12, …, 1/15}, the ht-index remains unchanged. This implies that ht-index is not sensitive to some small changes (Gao et al. 2016a, 2016b, 2017), although it has been used for characterizing fractal cities and hierarchical scaling (e.g., Long 2016). The ht-index, as previously defined, is likely to be rounded from a fractional ht-index (fht-index). In other words, scaling hierarchy could be a fraction. The present paper aims to seek a more precise ht-index – namely, the fht-index – for characterizing hierarchy of a data series with a heavy-tailed distribution. This paper further assigns an fht-index to an individual data value of a data series indicating its appropriate hierarchical level.

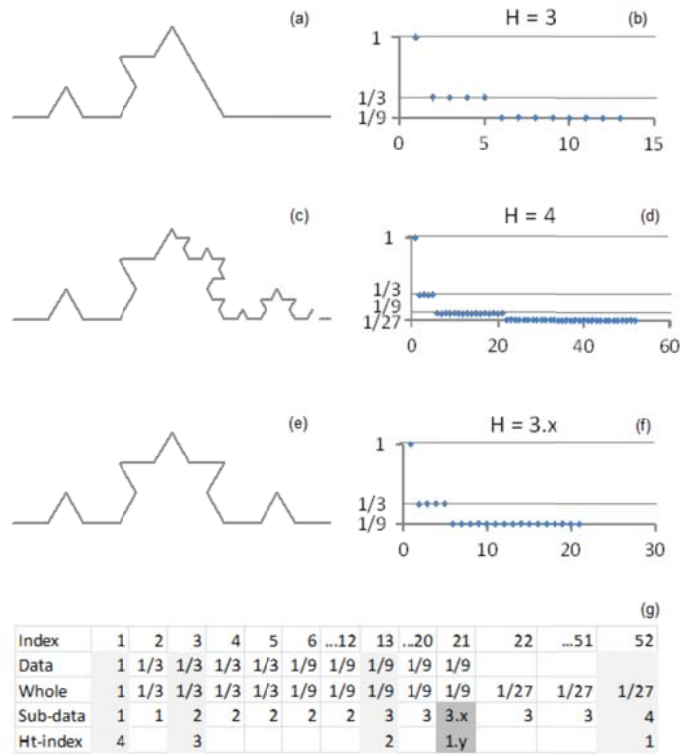

Figure 1. (Color online) Motivating and calculating the fht-index using the three different Koch curves (Note: Seen from the recursive perspective, there are 13 segments for the curve in panel (a) – one of scale 1, four of scale 1/3, and eight of scale 1/9 – so the notion of far more short segments than long ones recurs twice, as shown in panel (b). There are 52 segments for the curve in panel (c) – one of scale 1, four of scale 1/3, eight of scale 1/9, and 31 of scale 1/27 – so the scaling hierarchy is 4. There are 21 segments for the curve in panel (e) – one of scale 1, four of scale 1/3, 16 of scale 1/9 – but the plot in panel (f) shows only integral part of the ht-index of 3. The Koch curves in panels (e) and (c) are represented as data series and whole, respectively, in panel (g) for the purpose of computing fht-index. There are multiple sub-data series or sub-wholes with ht-indexes of 1, 2, and 3; these are also called anchors and constitute nested relationships with its whole.)

To further motivate the fht-index, let us examine three different Koch curves at different iterations of the generation (Figure 1). The two curves shown in panels (a) and (c) of Figure 1 consist of 13 and 52 segments, respectively, and their ht-indexes are 3 and 4, because the recurring times of far more short segments than long ones are 2 and 3 as shown in panels (b) and (d) of the same figure. These two ht-indexes are exactly 3 and 4, because removing one of the shortest segments from these two curves would not obtain the ht-indexes of 3 and 4, while adding one of the shortest segments would not increase the ht-indexes because of its insensitivity. From these two indexes, we can conclude that the Koch curve shown in panel (e) of Figure 1 must have an fht-index of 3.x, where 0 < x <1. However, ht-index as previously defined (Jiang and Yin 2014) captures only approximately scaling hierarchy, and is therefore less sensitive to some small changes. This is what motivates us to develop the fht-index.

### III. Wholes and sub-wholes

A fundamental concept of this paper is whole or sub-wholes. Assuming that the above ten numbers constitute a complete whole, the first three numbers or the first head would be a sub-whole. In other words, given a data series as a whole, its head and the head of the head (in a recursive fashion) would be the sub-wholes. This is just a simple understanding of whole or sub-wholes. The reader needs to refer to the following formal definition and methods for better understanding the whole or sub-wholes. It is important to realize that the curve shown in panel (e) of Figure 1 is not a whole, but part of a whole – the curve shown in panel (c) of the same figure. In this paper, a whole is defined as a data series of *n* values that ranges from the largest to smallest and meets the following condition: *ht-index(n) – ht-index(n-1) = 1*. For example, the 52 segments constitute a whole because *ht-index(52) – ht-index(51) = 1*. This definition of whole applies to sub-wholes as well. For example, the first 13 values of the 52 segments constitute a sub-whole because *ht-index(13) – ht-index(12) = 1*. According to the definition of whole or sub-whole, a Koch curve is not a whole, but the seemingly incomplete Koch curves shown in panels (a) and (c) are a sub-whole or a whole. In other words, the curve in panel (e) of Figure 1 is a whole according to the strict definition of Koch curve, but it is not a whole according to the very definition of head/tail breaks.

Given the 52 segments as a whole, ranking all its segments from the longest (of scale 1) to the shortest (of scale 1/27) creates a data series shown in panel (g) of Figure 1 – the row named "whole" – where data and its whole are shown together with its index in the first three rows. We have already derived the sub-whole of the 13 segments in the previous paragraph with the ht-index of 3. We further determine other sub-wholes or sub-data: the first three segments {1, 1/3, 1/3} with the ht-index of 2, and the first segment {1} with the ht-index of 1. All these sub-wholes (or sub-data series) are with integral ht-indexes as shown in panel (g) of Figure 1. These indexes with integral ht-indexes are called anchors for each sub-whole or whole. Note that the sub-whole and whole constitute a nested relationship; that is, the first sub-whole is within the second sub-whole, the first two sub-wholes are within the third sub-whole, and all the three sub-wholes are within the whole.

### IV. Methods – fht-index for a data series and its individual data

In order to determine the fht-index of the first 21 segments, we divided the data series range between the 13[th] and the 52[nd] (or the range between the third and fourth anchors) equally into 39 intervals and converted the equal intervals from a linear scale to a nonlinear scale using a power function of $f(interval_j) = (j * interval)^2$, where *j* is the index of each interval. This provides us with the fht-index of the first 21 segments: 3.042 (or x = 0.042 in panel (g) of Figure 1).

To summarize the calculation of the fht-index in general, given a data series, we first seek its whole by appending new data values up to the next hierarchical level, and sub-wholes by shrinking the data series to previous levels recursively. A whole is obtained from a data series by appending small values at its smallest end until the ht-index is increased to the next level exactly. In a similar vein, starting from the first value as the first sub-whole, more sub-wholes are obtained by adding values one by one until ht-index is increased to a next level exactly. A whole and its sub-wholes constitute nesting relationships. As a rule for determining the whole and sub-wholes or the anchors, the ht-index at index k must meet the condition of *ht-index(k) - ht-index(k-1) = 1*. Next, the range between two largest anchors, representing the largest sub-whole and the whole, respectively, should be equally interpolated and the equal intervals are then converted into a nonlinear scale to get the fht-index of a data series.

Having obtained the fht-index of the data series, we assign an fht-index to each data value of the data series. There are two ways to do this. The first is to take a whole whose ht-index is an integer, and the other is to take the data series (which is unlikely to be a whole) whose ht-index is a fraction. The data series to be examined is usually unlikely to be exactly a whole. Nevertheless, the input data series could incidentally be a whole. As shown in panel (g) of Figure 1, the largest data value is assigned to the first anchor, so it has the highest ht-index of 4, and the smallest data value is assigned to the fourth anchor, so it has the lowest ht-index of 1. Having assigned all integral ht-indexes to these anchors, other indexes are assigned to some fht-indexes by interpolating the ranges between these anchors. This assignment of integral ht-indexes looks like the flip process of determining anchors; the anchors increase from the first data value to the last, while the integral ht-indexes decrease from the first data value to the last. After assigning the integral ht-indexes, we have to interpolate the range between sub-wholes and the range between sub-whole and the whole in order to obtain fht-indexes of other individual values. Eventually, the fht-index of the 21$^{st}$ segment is 1.63 (or y = 0.63 in panel (g) of Figure 1). The above procedure for a whole can be packed as a function of the fht-index:

```
Function Fht-index (whole)
     // This function returns a fht-index for each value in the whole
     // The data is sorted ranging from the largest to the smallest
     Anchors (whole)
     // this function returns AnchorNum of the whole
     Flip AnchorNum in whole;
     // The largest AnchorNum is assigned to the lowest marked index, while
     //the smallest AnchorNum is assigned to the largest marked index
     Foreach marked index p:
         Find its next marked index p';
         range = p' - p;
         subHtFraction = Interpolation (AnchorNum, range);
         htFraction.add(subHtFraction);
     Return htFraction;
End Function
```

For a data series that is not incidentally a whole, it is necessary to append some smallest values in order to make it a whole. While this is simple for the Koch curves, for real-world data it is important to get its trend line that best fits the data series. In this regard, it is recommended to use trend line functions such as power law, logarithmic, polynominal, and exponential. As a rule, the most-fit trend line must be chosen for a specific data series. The fht-index (e.g., 3.x) of the data series is obtained by interpolating the range between the largest sub-whole and the whole. The anchors are with integral ht-indexes, but in the opposite order; the largest anchors with the smallest integral ht-index, and smallest anchors with the largest integral ht-index. Those data values between anchors or between the largest anchors and the whole must be obtained through interpolation. To this point, we have relied on the Koch curves to illustrate the ideas of fht-index in order to make it more accessible to experts as well as non-experts.

**V. Case studies and FHTCalculator**
To further explore the fht-index, we applied it to two case studies. The first case study involves 36 city sizes that follow Zipf's law (Zipf 1949) exactly: 1, 1/2, 1/3,…, and 1/36 (panel (a) of Figure 2) with an ht-index of 3. The second case study involves 8,106 natural cities with an ht-index of 7, derived from the social media Brightkite in the United States (panel (c) of Figure 2, Jiang and Miao 2015). For the first case study, appending the smallest values is pre-determined by the rank sizes, while for the second case study the smallest values are determined by a power law function of $P = 5,03\, r^{-2.1}$ of the 8,016 city sizes. Unlike the ht-indexes that are discrete, fht-indexes for individual data values, as shown in panels (b) and (d), are continuous, and thus capture scaling hierarchy more precisely than the discrete ht-indexes. The fht-indexes of these two data are 3.81 and 7.04, respectively based on the methods introduced above, or by applying these data series into FHTCalculator (2017). The fht-index of individual value within the two data is plotted in Figure 2. Note that the fht-indexes are not simply interpolated from the discrete ht-index, but are recalculated from their wholes as described above. We developed a small program for computing fht-index, called FHTCalculator (2017). The computing for the two case studies can be done within a few seconds. This program has been made available in GitHub, and interested readers can try it with their own data.

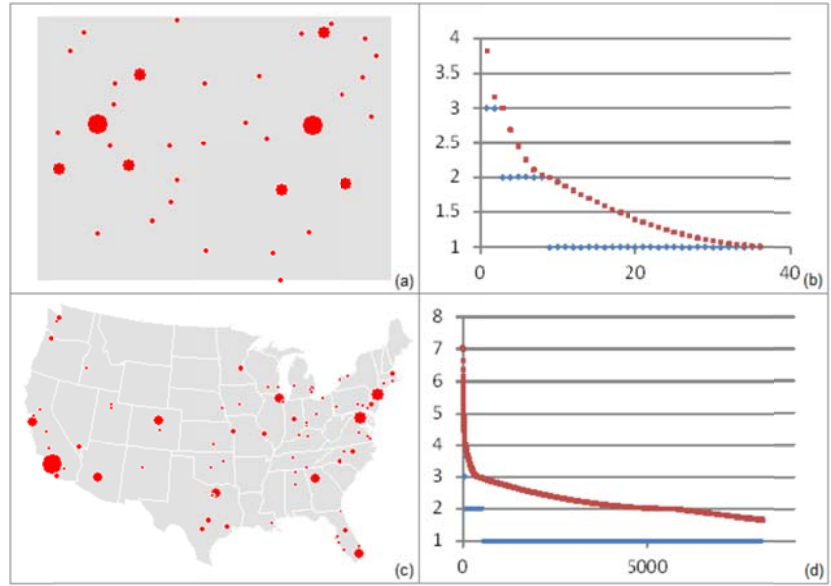

Figure 2. (Color online) Two case studies for visualization of fractional hierarchy
(Note: The 36 city sizes that follow Zipf's law precisely (a); The ht-index in blue and fht-index in red of the 36 city sizes (b); Top five hierarchical levels of 8106 natural cities in the USA (c); The ht-index in blue and fht-index in red of the 8106 natural cities (d). It should be noted that scaling hierarchies with the two data sets are more precisely analyzed and visualized by the fht-index than the ht-index initially. This fact is clearly seen in panels (b) and (d).)

## VI. Implications

Existing fractals, both classic and statistical, are essentially defined from the top down, i.e., either a strict or statistical fractal can be generated by following a rule endlessly, such as the Koch curve or the statistical Koch curve; see the literature on the theory (Mandelbrot 1967, 1982) and its applications in geography (e.g., Batty and Longley 1994, Frankhauser 1994, Chen 2011). The new relaxed definition or the third definition of fractal is imposed from the bottom up, capturing the underlying scaling hierarchy of far more small things than large ones through the ht-index. The fht-index takes a step further because it can more precisely measure the degree of hierarchy from a previously discrete value to a continuous value, or from a previous integer to a real number.

This continuous value is sensitive enough to capture different phrases of a fractal from its initial stage to matured stage with the fht-index increasing slowly or gradually (rather than rapidly as the ht-index). It is in this sense that we believe that the new fractal geometry with a focus not only on statics but also on dynamics. This new fractal geometry is very much in line with living geometry developed by Alexander (2002-2005). The living geometry aims not only for understanding fractal structure but also for making complex or living structure. In this connection, the fht-index provides an excellent means for judging living structure, for example, when it is applied for measuring degree of livingness (Jiang 2015b). In addition, the fht-index can more precisely characterize spatial heterogeneity that is more pervasive or ubiquitous in geography (Jiang 2015c). By focusing on fractal structure of far more small things than large ones, in addition to spatially auto-correlated things (Tobler 1970), the fht-index represents a new approach for geospatial analysis.

## VII. Conclusion

This paper refines the ht-index to be a fraction to better characterize the scaling hierarchy of a fractal or data series with a heavy-tailed distribution. The existing integral ht-index is implicitly based on the assumption that any given data series of a heavy-tailed distribution is always a whole. This assumption does not always hold true. In many cases, a data series is likely to be part of a whole rather than a whole itself. Based on this new perception, we put a data series within a whole and seek its sub-wholes or anchors in order to derive its fht-index. This fht-index is always greater than the integral ht-index. We further assign an fht-index to each data value of the data series. More precisely, the anchors are with integral ht-indexes, while other data values or non-anchors are with fht-indexes. The fht-index may help measure degree of living structure or more efficiently and effectively visualize fractal urban structure and nonlinear dynamics, since the structure and dynamics have been firstly captured by the fht-index. In the future, we will seek applications of the fht-index to better characterize geographic forms and processes, or urban structure and dynamics in particular, and even beyond the understanding towards the making – how to better heal and design built environments.

**Acknowledgement**
XXXXXXX